\documentclass{article}
\usepackage{amsmath,mathrsfs,amssymb} 
\usepackage{graphicx,hyperref}

\newcommand{\papertitle}{%
Perfect transmission scattering as a $\mathcal{PT}$-symmetric spectral problem}
\newcommand{\pauthor}{%
H. Hernandez-Coronado$^{a}$, D. Krej\v{c}i\v{r}\'{i}k$^{b,c,d}$ and P. Siegl$^{a,b,e}$}
\begin{document}

\vspace*{-1cm}
\begin{flushright}
\textsf{}
\\
\mbox{}
\\

\end{flushright}
\begin{LARGE}
\bfseries{\papertitle}
\end{LARGE}
\vspace{1cm}

\begin{large}%
\pauthor
\end{large}

\vspace*{.1cm}

\phantom{XX}
\begin{minipage}{1\textwidth}
\begin{flushleft}
\begin{it}
$^{a}$ FNSPE, Czech Technical University in Prague, 11519 Prague, Czech Republic.\\
$^{b}$ Department of Theoretical Physics, Nuclear Physical Institute ASCR, 25068 \v{R}e\v{z}, 
Czech Republic.\\
$^{c}$ Basque Center for Applied Mathematics, Bizkaia Technology Park 500, 48160 Derio, Spain.\\
$^{d}$ IKERBASQUE, Basque Foundation for Science, 48011 Bilbao, Spain.\\
$^{e}$ Laboratoire Astroparticule et Cosmologie, Universit\'e Paris 7, 75205 Paris, France.

\end{it}
\end{flushleft}
\texttt{\footnotemark[1]coronhec@fjfi.cvut.cz,  krejcirik@ujf.cas.cz, siegl@ujf.cas.cz
\phantom{X}}
\end{minipage}

\vspace*{1cm}
\begin{abstract}
 We establish that a perfect-transmission scattering problem can be described by a class of parity and time reversal symmetric 
operators and hereby we provide a scenario for understanding and implementing the corresponding quasi-Hermitian quantum 
mechanical framework from the physical viewpoint. 
One of the most interesting features of the analysis is that the complex eigenvalues of the underlying non-Hermitian 
problem, associated with a reflectionless scattering system, lead to the loss of perfect-transmission energies as the parameters 
characterizing the scattering potential are varied. On the other hand, the scattering data can serve to describe 
the spectrum of a large class of Schr\"odinger operators with complex Robin boundary conditions. 
\end{abstract}
\vspace*{.5cm}
\vspace*{.5cm}

\section{Introduction}

Recently there has been a considerable amount of work devoted to the study of the so-called 
$\mathcal{PT}$-symmetric quantum mechanics -- see \cite{Bender-2007-70, Mostafazadeh-2010-review} and the references therein.
The theory is based on the idea to give a physical meaning to a class of non-Hermitian Hamiltonians being symmetric under the composed space reversal transformation $\mathcal{P}$ and complex conjugation $\mathcal{T}$. These Hamiltonians are interesting because some of them have exclusively real spectrum and -- usually only after an appropriate change of the inner product of Hilbert space -- they can generate unitary time-evolution. The relevance of $\mathcal{PT}$-symmetric operators has been suggested in various domains of physics: nuclear physics \cite{Scholtz-1992-213}, optics \cite{Klaiman-2008-101,Longhi-2010-105,Schomerus-2010-104,West-2010-104}, solid state \cite{Bendix-2009-103}, superconductivity \cite{Rubinstein-2007-99,Rubinstein-2010-195}, and electromagnetism 
\cite{Ruschhaupt-2005-38,Mostafazadeh-2009-102}. Moreover, the first experimental results in optics using the formalism of $\mathcal{PT}$-symmetric quantum mechanics in the theoretical explanation of observed effects have appeared recently \cite{Longhi-2009-103,Guo-2009-103,Ruter-2010-6}.

In this letter, we establish a purely quantum-mechanical interpretation of a class of $\mathcal{PT}$-symmetric
Hamiltonians in a particular scattering problem. Hereby we confirm the common claim that 
the non-Hermiticity corresponds to gain/loss mechanisms of probability density and that the presence 
$\mathcal{PT}$-symmetry ensures the balance between these opposite effects  \cite{West-2010-104,Bendix-2009-103}.    
We show that the spectral techniques for non-self-adjoint (NSA) problems can be used for describing a scattering system
in the reflectionless regime.

The general idea is that the above kind of scattering problems can be described by an effective
 Schr\"odinger equation in a bounded interval (corresponding to the domain of the scatterer) subject to complex 
Robin boundary conditions. This problem can be regarded as a particular class of $\mathcal{PT}$-symmetric 
quantum problems if the scattering potential respects such a symmetry. The gain/loss mechanism referred to above is clearly understood in this case, since we start with a Hermitian physical system and we can keep track of where and how the probability is lost or gained. 
Furthermore, the typical complex points appearing in the spectra of the NSA problems have a very natural explanation: they give rise to the loss of the perfect-transmission energies (PTEs). Finally, we solve the inverse problem, \emph{i.e.} how 
the spectrum of a $\mathcal{PT}$-symmetric problem can be determined from the knowledge of PTEs. 

We will conclude with the statement that the real points in the spectra of certain class of $\mathcal{PT}$-symmetric 
Hamiltonians can be \emph{measured} in the quantum mechanical scattering problem and the points where two eigenvalues coalesce (exceptional points) correspond to the loss of PTEs.

\section{From scattering to spectral theory}\label{sec:SST}

Consider a quantum particle of mass $m$ scattered by a potential of the form $V(x,y,z)=v(x)$, for a general real-valued function $v$ supported in $[-a,a]$ with $a>0$. We shall restrict ourselves to scattering waves in the $x$-direction, so that the problem can be described by the one-dimensional Schr\"odinger equation
\begin{equation}
 -\psi''(x)+v(v)\psi(x)=k^2\psi(x),\label{sch1deq}
\end{equation}

where $\psi$ is the particle wavefunction and $k$ a positive (wave)number. Since~$v$ is zero outside $[-a,a]$, we have the asymptotic solutions: 
$\psi_l(x)=\exp{(ikx)} + R \exp{(-ikx)}$ 
for the in-coming wave (for $x\le -a$), and $\psi_r(x)=T \exp{(ikx)}$ for the out-coming wave (for $x\ge a$), 
where $R$ and $T$ correspond to the reflection and transmission amplitudes,
respectively. Note that we consider only the special solutions
for which the incident amplitude is equal to one.

Now we explain how the state $\psi$ of the particle can be described by an effective Schr\"odinger equation 
in the interval $[-a,a]$.
We focus on the special case of perfect-transmission, {\it i.e.}~$R=0$. By plugging $R=0$ in the in-coming
wave and requiring the continuity of~$\psi$ and its derivative at the boundary~$\pm a$, it is easy to show that 
the scattering problem is then equivalent to the non-linear (energy dependent) problem \mbox{(in units where $m=1/2$ and $\hbar=1$)}
\begin{eqnarray}
 -\psi''(x) + v(x) \psi(x)&=&k^2 \psi(x),\qquad
\forall x\in[-a,a],\label{s-nonlinear}\\
 \psi'(\pm a) - i k \psi(\pm a)&=&0.\label{rbcfull}
\end{eqnarray}%

The non-linear problem \eqref{s-nonlinear}--\eqref{rbcfull} can be solved by considering the associated one-parametric (linear) spectral problem:
\begin{eqnarray}
 -\psi''(x) + v(x) \psi(x)&=&\mu(\alpha) \psi(x),\quad
\forall x\in[-a,a],\label{mu-nonlinear}\\
 \psi'(\pm a) - i \alpha\psi(\pm a)&=&0.\qquad \label{rbcalpha}
\end{eqnarray}%
In the above expression, ~$\mu(\alpha)$ plays the role of eigenvalue and $\alpha$ is a real parameter. The energies 
corresponding to the perfect-transmission states are found as those points $\mu(\alpha_*)$ satisfying 
\begin{equation}\label{s-intersection}
\mu(\alpha_*) = \alpha_*^2.
\end{equation}

The relation \eqref{rbcalpha} is the so-called $\mathcal{PT}$-symmetric Robin boundary condition and 
it has been studied before in the context of spectral theory for NSA operators 
\cite{Krejcirik-2006-39, Krejcirik-2008-41a, Krejcirik-2010}. These boundary conditions have been used previously in the phenomenological
description of emission and absorption, however, here they appear naturally in the scattering 
and in the framework of spectral theory for $\mathcal{PT}$-symmetric (or more general NSA) operators. Moreover, since the boundary conditions (\ref{rbcalpha}) imply that the probability current at $x=\pm a$ does not vanish (for $\alpha\neq 0$), the non-self-adjointness in this system can be clearly understood as the gain/loss of probability density at the boundary points. Furthermore, the $\mathcal{PT}$ symmetry is a consequence of our restriction to the reflectionless regime, which corresponds to the exact compensation of the gains and losses.

\section{Square well potential}
We illustrate our approach on the explicitly solvable model of the square well 
\begin{equation}
 v(x)=-v_0 \chi_{[-a,a]}(x),\label{vsw}
\end{equation}
where $\chi_{A}(x)$ is the characteristic (or indicator) function of a set $A$ and $v_0>0$. The eigenvalues of the corresponding NSA problem \eqref{mu-nonlinear}--\eqref{rbcalpha} with the above potential are given by (see \cite{Krejcirik-2006-39}):
\begin{equation}
 \mu_n =
\left\{
\begin{array}{l l}
 \alpha^2-v_0, & n=0,\\
(\frac{n\pi}{2a})^2-v_0, & n \in \mathbb{N}.
\end{array}
\right.\label{evaw}
\end{equation}

Therefore, by employing the knowledge of the spectrum of the Hamiltonian and taking into account expression
 \eqref{s-intersection}, we get indeed the well-known PTEs for the square well potential 
$k^2 = (\frac{n\pi}{2a})^2 - v_0$ 
(see {\it e.g.}, \cite{Cohen-Tannoudji-1977-1}). 
\begin{figure}[htb!]
 \centering
 \includegraphics[width=0.55\textwidth]{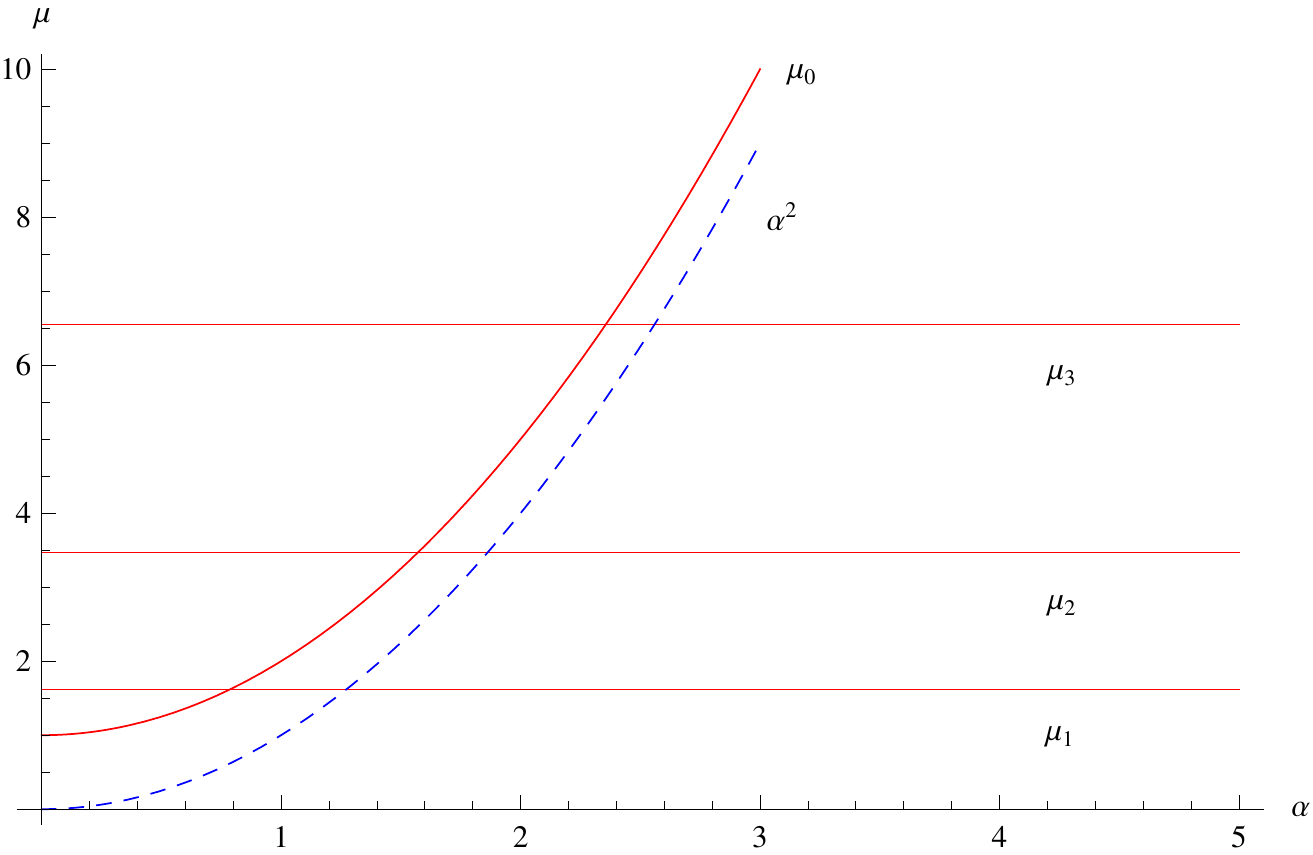}
 \caption{(Color online) 
The spectrum of the square well potential
with $v_0=-1$ and $a=2$. The dashed (blue) curve represents the dispersion relation
curve $\mu=\alpha^2$ while the continuous (red) lines
are the  energy levels. The PTEs correspond to the intersections
of the constant $\mu_n$'s with the dispersion parabola.
}
 \label{fig1}
\end{figure}

The first three PTEs, corresponding to the intersection of the graphics $\mu_n(\alpha)$ and expression \eqref{s-intersection}, are presented in Figure \ref{fig1}. If the potential depth $v_0$ tends to zero, the continuous (red) parabola corresponding to the eigenvalue $\mu_0$ approaches the dashed (blue) parabola representing dispersion relation 
\eqref{s-intersection}. The two parabolas coincide when $v_0=0$, \emph{i.e.}\ with no potential all positive energies trivially correspond to perfect transmission.

\section{Multiple steps potential}

We claim that the presence of complex eigenvalues lead to observable effects. We demonstrate this
on models with even piecewise constant potentials 
\begin{equation}
v(x)=\sum_{j=1}^{N+1} \beta_j \varepsilon_j^{-1} \chi_{[x_{j-1},x_j]}(|x|),\label{vms}
\end{equation}
where 
$0 \leq x_0 < \dots < x_N \leq a$, $\varepsilon_j=x_j-x_{j-1}$ 
determine the width and $\beta_j$ the strength of the constant parts. This type of solvable models -- as usual by using the 
explicitly known wavefunctions in the intervals where the potential is constant and matching them at the interface points -- 
can serve as approximation to realistic physical potentials which also fit to our framework. 

\begin{figure}[htb!]
 \centering
 \includegraphics[width=0.55\textwidth]{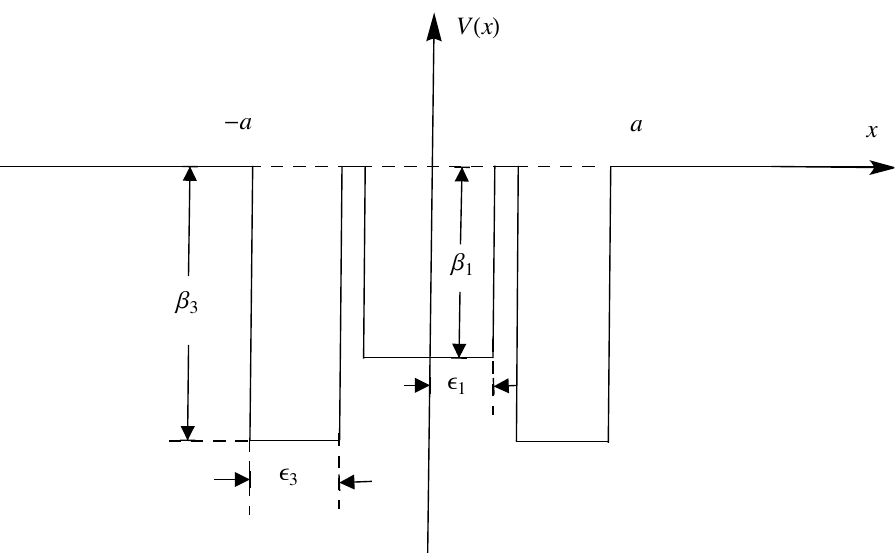}
 \caption{The shape of considered multiple steps potential.
}
 \label{fig5}
\end{figure}

\begin{figure}[htb!]
\centering
\includegraphics[width=0.55\textwidth]{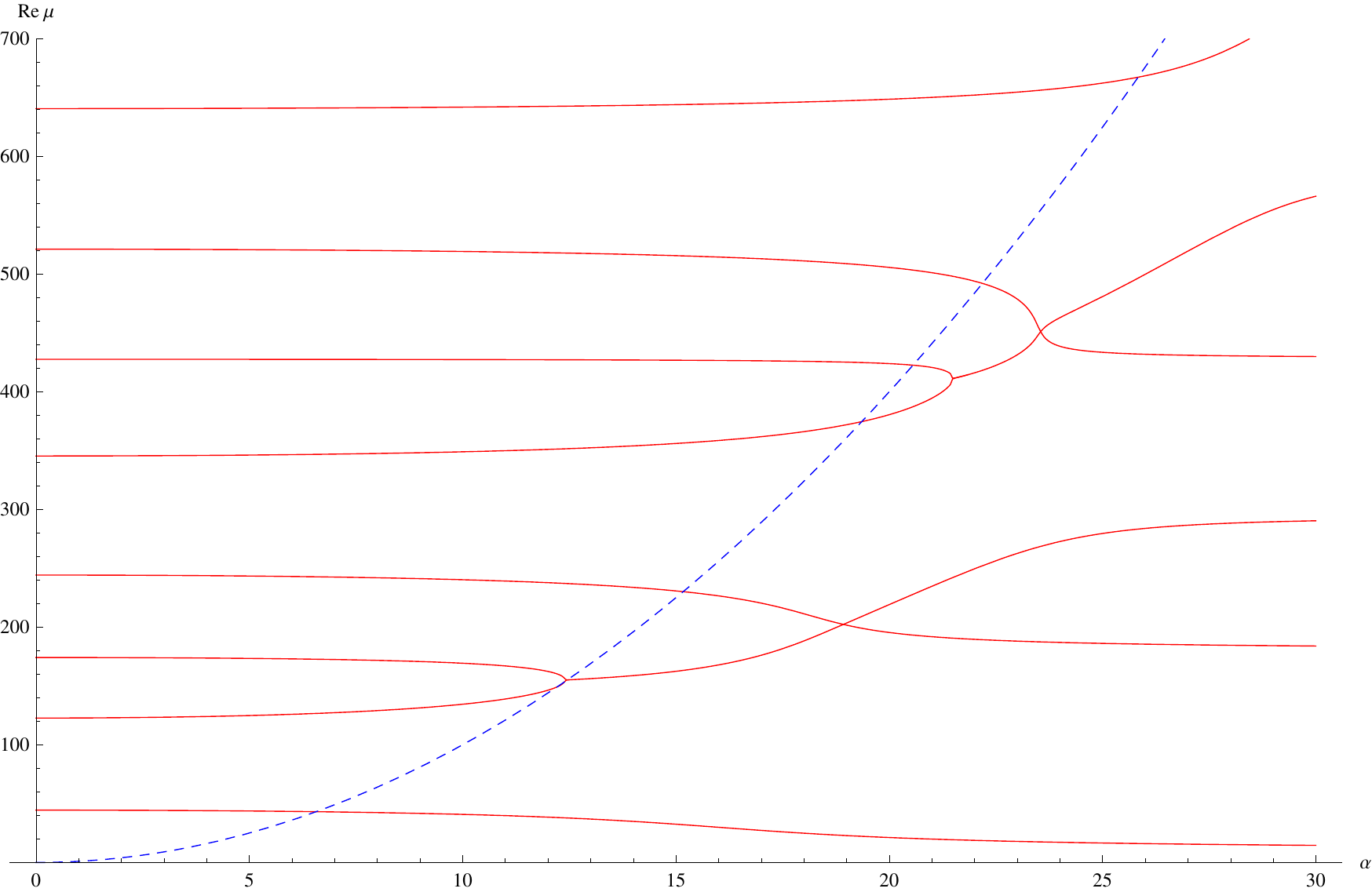}
\caption{(Color online) The real  part of the eigenvalue $\mu$ as a function of the parameter $\alpha$ for the step-like
potential $v$ with $a=\pi/4$, $\varepsilon_1=0.2$, $\varepsilon_3=0.5$, $\beta_1=-90$, $\beta_2=0$, $\beta_3=-100$. The PTEs correspond to the intersections of the energy levels (continuous, red) and the dispersion relation (dashed, blue).}
\label{fig2}
\end{figure}

We focus on $N=2$ model with $x_0=0, x_3=a$ and $\beta_1 < 0$, $\beta_2=0$, $\beta_3 \neq 0$, \emph{i.e.} two steps
localized at the endpoints and one at the origin, see Figure \ref{fig5}, however, our reasoning is not limited to this particular solvable 
potential. Inspired by the delta-interaction models \cite{Krejcirik-2010} that are limits of the considered potentials for a special choice of parameters, it is not surprising that complex conjugated pairs of eigenvalues are present in the spectrum. Intersections of the dispersion parabola \eqref{s-intersection} and energy levels represent PTEs, \emph{cf.}\ Figure \ref{fig2}. However, there is no PTE corresponding to the 
intersection with complex energy since $\mu=k^2$ is required to be real (to have in-coming and out-coming 
plane waves $e^{i k x}$). 

\begin{figure}[htb!]
\centering
\includegraphics[width=0.55\textwidth]{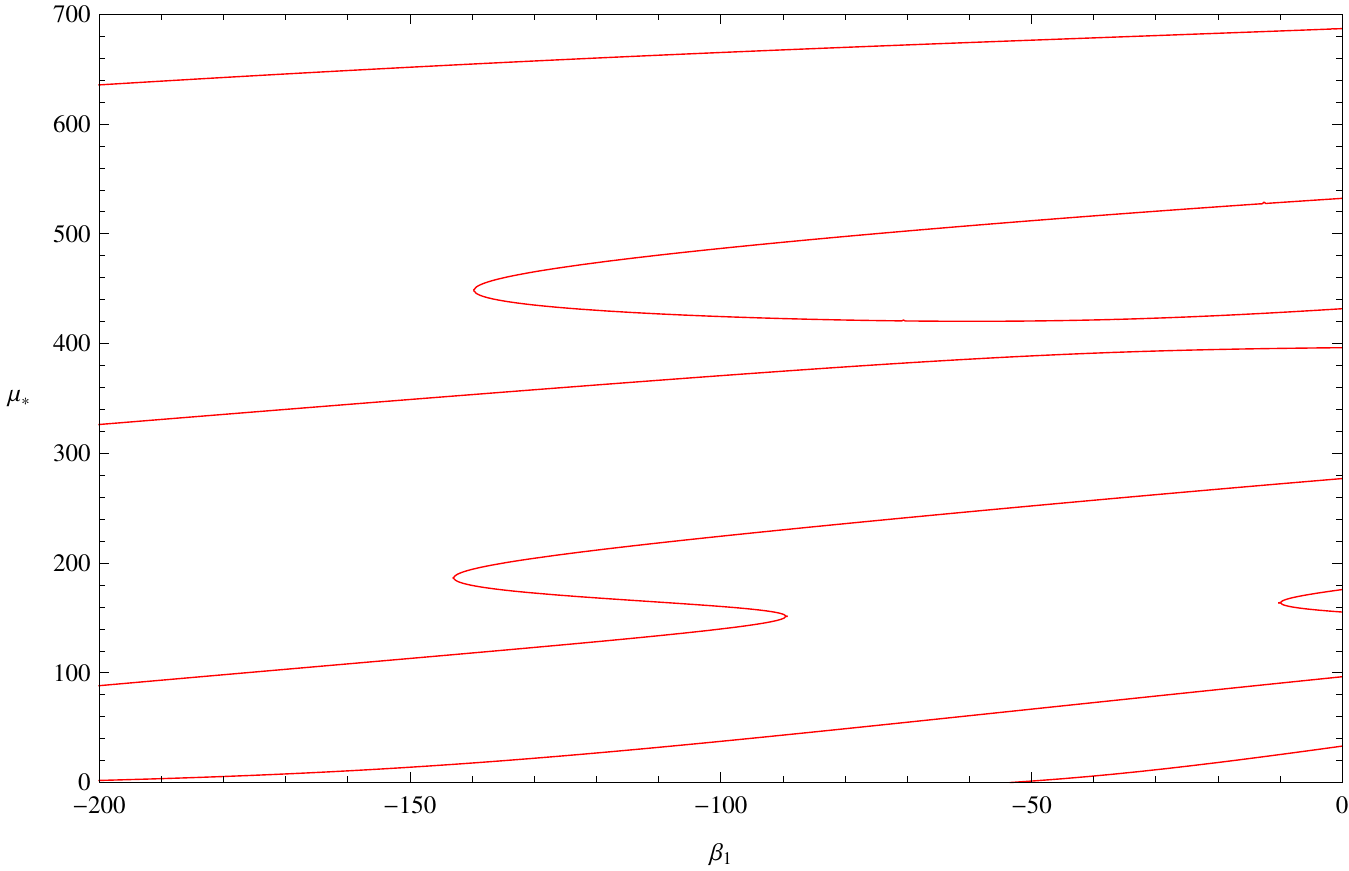}
\caption{(Color online) The PTEs $\mu_*$ as a function of $\beta_1$ for the step-like
potential $v$ with $a=\pi/4$, $\varepsilon_1=0.2$, $\varepsilon_3=0.5$, $\beta_3=-100$, and $\beta_2=0$. 
The losses of PTEs are clearly visible \emph{e.g.} around $\beta_1 =-140$, for $\mu_*\approx 190$ and $\mu_* \approx 450$.
}
\label{fig3}
\end{figure}

The shape of the energy curves $\mu_n(\alpha)$ depends on the potential and we prepare such scenario 
(by fixing the steps at the endpoints and changing the strength of the one in the middle) that the dispersion parabola 
intersects at first two energy levels, then precisely the point of complexification (or exceptional point), 
and finally the complex level. Figure \ref{fig3} illustrates the resulting behavior of PTEs, two originally separated PTEs 
merge when the intersection is the complexification point and then completely disappear, see animation \cite{Siegl-2010-PTE}.

\begin{figure}[htb!]
\centering
\includegraphics[width=0.55\textwidth]{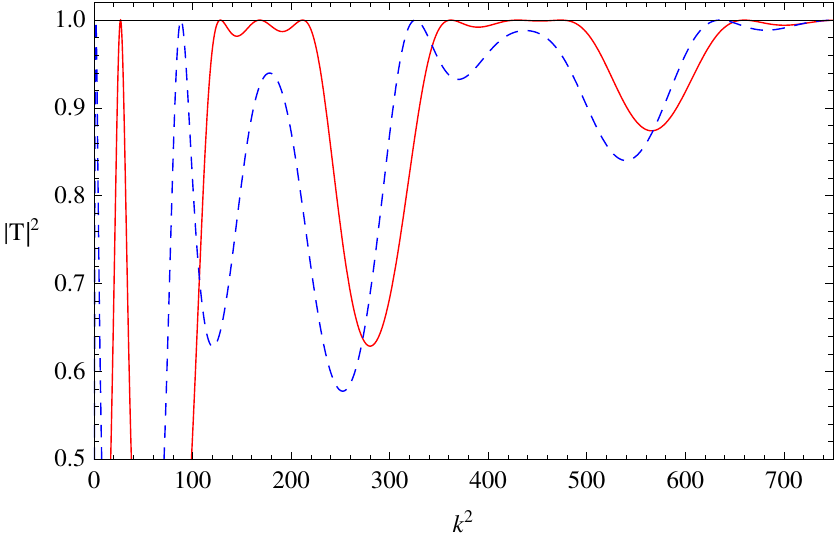}
\caption{(Color online) Transmissions $|T|^2$ as a function of energy $k^2$ for the step-like
potential $v$ with $a=\pi/4$, $\varepsilon_1=0.2$, $\varepsilon_3=0.5$, $\beta_3=-100$, $\beta_2=0$, $\beta_1=-120$  
(continuous red line), and $\beta_1=-200$ (dashed blue line). See \cite{Siegl-2010-PTE} for animated plots of $|T|^2$ as a function of potential. 
}
\label{fig4}
\end{figure}

This process can be also described in terms of the transmission
coefficient. The comparison of the two curves in Figure \ref{fig4}
(corresponding to different values of $\beta_1$) indicate 
how the process can be observed from scattering data: two initially
separated peaks (PTEs), {\it e.g.} around $k^2\approx 160$ in the
continuous red line, collide as $\beta_1$ is varied (complexification
point is intersected by the dispersion parabola) and fall down
afterwards (a loss of PTEs) around $k^2\approx 180$ in the dashed blue
line. Broad peaks close to $|T|^2=1$, \emph{e.g.} around $k^2=460$,
correspond to the collision of PTEs (approaching the complexification
point).  

\section{Inverse problem}
Now let us explore the inverse problem, \emph{i.e.} how we can determine the spectrum of a given $\mathcal{PT}$-symmetric
 (or more general NSA) Hamiltonian subject to complex Robin boundary conditions by measuring the PTEs in
 a scattering experiment.
Consider the NSA problem defined by expressions \eqref{mu-nonlinear}--\eqref{rbcalpha} together with the dispersion relation
\eqref{s-intersection} 
and let us modify it by adding a constant $v_0$ to the potential, \emph{i.e.} putting the system into a square well, as follows: 
\begin{eqnarray}
-\psi''(x)+(v(x)+v_0) \psi(x)&=&\mu_0(\alpha) \psi(x), \label{Schr1a}  \\
\psi'(\pm a) -i \alpha \psi(\pm a)&=&0,  \label{Schr1b}  \\
\mu_0(\alpha)&=&\alpha^2. \label{par1}
\end{eqnarray}
The above problem can be recast as
\begin{eqnarray}
-\psi''(x)+v(x) \psi(x) &= &\mu(\alpha) \psi(x), \label{Schr3a}\\
\psi'(\pm a) -i \alpha \psi(\pm a)&= &0,    \label{Schr3b} \\
\mu(\alpha)&=&\alpha^2-v_0, \label{par3}
\end{eqnarray}
where  we have introduced $\mu(\alpha):=\mu_0(\alpha)-v_0$. Clearly, (\ref{Schr3a})--(\ref{Schr3b}) is 
\emph{the same spectral problem} as the initial one (\ref{mu-nonlinear})--(\ref{rbcalpha}), however, by comparing 
expressions (\ref{s-intersection}) and (\ref{par3}), we see that the dispersion parabola is shifted by $v_0$. 

Now, consider measurements of the PTEs corresponding to the choice of $v_0$, 
which we assume to be discrete for every $v_0$. Let us choose a PTE and denote by
 $\kappa(v_0)$ its $v_0$-dependence. Thus we have a function $v_0 \mapsto \kappa(v_0)$. It is clear from
 \eqref{Schr1a}--\eqref{par3} that $\kappa(v_0)-v_0$ is the eigenvalue for the spectral problem
 \eqref{Schr3a}--\eqref{Schr3b} with $\alpha$ satisfying $\alpha^2=\kappa(v_0)$. Hence we can obtain the entire
 $\alpha$-dependence of the chosen energy level $\mu(\alpha)$ if $\kappa$ is an invertible function. 
Indeed, for every $\alpha$, we find $v_0=\kappa^{-1}(\alpha^2)$ such that 
$\mu(\alpha)=
\alpha^2-\kappa^{-1}(\alpha^2)$
is the eigenvalue of \eqref{Schr3a}--\eqref{Schr3b}.

It remains to ensure that $\kappa$ is invertible, which follows if $\kappa'(v_0) \neq 0 $ for all $v_0$. 
For all $v_0$, we have 
$\kappa(v_0)=\mu(\alpha)+v_0,$
with $\alpha^2=v_0$. By differentiating this relation with respect to $v_0$, we obtain
$\kappa'(v_0) =2 \alpha / (2\alpha -  \mu'(\alpha))$.
The situation $2\alpha = \mu'(\alpha)$ is very particular: $\mu(\alpha)$ is (at least) locally parabola, thus
it represents either reflectionless setting
(the dispersion parabola \eqref{par3} locally coincides with the energy level $\mu(\alpha)$) or no perfect transmission
(no intersection of the energy level $\mu(\alpha)$ and (\ref{par3})). 
Besides these exceptional cases we have obtained 
the condition on the spectrum of 
\eqref{Schr3a}--\eqref{Schr3b} 
$|\mu'(\alpha)|<\infty$
that assures invertibility of $\kappa$. This condition is however satisfied for all real $\alpha$, except possibly for the points where energy levels cross, because of the analyticity of $\mu(\alpha)$, which is ensured for all relatively form bounded potentials $v$.  

This means that we can determine the spectrum of  the problem (\ref{Schr3a})--(\ref{Schr3b}) by \emph{measuring} the
PTEs as a function of $v_0$. Of course, by this procedure we can only determine the real
eigenvalues of the corresponding spectrum. The appearance of a complex conjugate pair of eigenvalues can be traced
back as a loss of two (close) PTEs, analogously for the restoration of two real eigenvalues. 

\section{Discussion and conclusions}
We have shown that NSA Hamiltonians naturally arise in the effective description of a class of scattering problems. Within this context, we have proposed a spectral-type scheme for obtaining the energies corresponding to a perfect-transmission scattering process. The model confirms the common claim that $\mathcal{PT}$-symmetric operators describe physical systems where the probability density is not conserved locally, but the gains and losses compensate globally. The appearance of complex eigenvalues in the associated $\mathcal{PT}$-symmetric spectral problem -- resulting from the collision of a pair of real ones -- can lead to the merging and subsequent disappearance of two PTEs which, moreover, can be observed in a scattering experiment. Furthermore, we have discussed the inverse problem, that is, how to use the scattering data to determine the spectrum of an operator with non-Hermitian boundary conditions.

It is important to stress that the general idea of reducing a scattering problem to a non-linear eigenvalue equation
is not limited to even potentials (ensuring the $\mathcal{PT}$-symmetry) and the $\mathcal{PT}$-symmetric Robin-type boundary conditions (reflecting the perfect-transmission process). However, when $\mathcal{PT}$-symmetry is relaxed, real energy
levels do not need to cross anymore to produce complex eigenvalues and therefore the overall 
analysis of the spectrum of the associated NSA operator is more complicated. 

The main purpose of this letter was to establish a truly quantum-mechanical interpretation of a $\mathcal{PT}$-symmetric model. Furthermore, we believe that the associated spectral framework of perfect-transmission process provides an additional insight to effects that can be observed in scattering data.

\section*{Acknowledgments}

This work has been partially supported by the Czech Ministry of Education, Youth and Sports within the project LC06002. P.S.\ appreciates the support by the Grant Agency of the Czech Republic project No.\ 202/08/H072 and by the Grant Agency of the Czech Technical University in Prague, grant No.\ SGS OHK4-010/10.  

{
\footnotesize
\bibliographystyle{abbrv}
\bibliography{references} }

\end{document}